\documentclass[%
 amsmath,amssymb,
 aps,
 prl,
twocolumn,
superscriptaddress
]{revtex4-2}
\pdfoutput=1

\usepackage[dvipsnames]{xcolor}
\usepackage{graphicx}
\usepackage{dcolumn}
\usepackage{bm}
\usepackage[colorlinks=true,linkcolor=blue, citecolor=blue, urlcolor=blue, bookmarks]{hyperref}
\usepackage{braket}
\usepackage{ulem}
\usepackage{lineno}
\usepackage{soul}

 \newcommand{\review}[1]{{#1}}

 \renewcommand{\sout}[1]{}
\begin{document}

\title{Detecting a long lived false vacuum with quantum quenches}

\author{Gianluca Lagnese}
\affiliation{Institute of Polar Sciences CNR, Via Torino 155, 30172 Mestre-Venezia, Italy}
\affiliation{SISSA, via Bonomea 265, 34136 Trieste, Italy}
\affiliation{Jo\v zef Stefan Institute, 1000 Ljubljana, Slovenija}
\author{Federica Maria Surace}
\affiliation{Department of Physics and Institute for Quantum Information and Matter,
California Institute of Technology, Pasadena, California 91125, USA}
\author{Sid Morampudi}
\affiliation{Center for Theoretical Physics, Massachusetts
Institute of Technology, Cambridge, MA 02139, USA}
\author{Frank Wilczek}
\affiliation{Center for Theoretical Physics, Massachusetts
Institute of Technology, Cambridge, MA 02139, USA}
\affiliation{T. D. Lee Institute and Wilczek Quantum Center,
SJTU, Shanghai 200240, China}
\affiliation{Arizona State University,
Tempe, AZ 25287, USA}
\affiliation{Stockholm University,
Stockholm 10691, Sweden}

\date{\today}

\begin{abstract}
Distinguishing whether a system supports alternate low-energy (locally stable) states -- stable (true vacuum) versus metastable (false vacuum) -- by direct observation can be difficult when the lifetime of the state is very long but otherwise unknown.
Here we demonstrate, in a tractable model system, that there are physical phenomena on much shorter time scales that can diagnose the difference. 
Specifically, we study the time evolution of the magnetization following a quench in the tilted quantum Ising model, and show that its magnitude spectrum is an effective diagnostic.
Small transition bubbles are more common than large ones, and we see characteristic differences in the size dependence of bubble lifetimes even well below the critical size for false vacuum decay.  We expect this sort of behavior to be generic in systems of this kind. We show such signatures persist in a continuum field theory. This also opens the possibility of similar signatures of the potential metastable false vacuum of our universe well before the beginning of a decay process to the true vacuum.
\end{abstract}

\maketitle

Many interesting physical systems, possibly including our present-day universe \cite{ELIASMIRO2012222}, can exist in metastable states. The decay of a metastable state (or {\it false vacuum}) is in general understood as a phenomenon of bubble nucleation \cite{LANGER1967108, Kobzarev:1974cp}.
Nucleation is a general phenomenon of first order phase transitions, where the new phase is not reached uniformly across the whole system, but rather through the formation of finite domains ({\it bubbles}).
The theory of thermal bubble nucleation was introduced in Langer's pionering work \cite{LANGER1967108} on Ising ferromagnets and extended to quantum field theory by Kobzarev \textit{et al.} \cite{Kobzarev:1974cp}.  The quantum version was deeply analyzed in seminal work by Sidney Coleman 
\cite{coleman_1977}.  In a quantum system, bubbles of true vacuum arise as quantum fluctuations. Most bubbles have a large energy cost associated with the surface tension of their walls, and are therefore transient, or virtual. Bubbles larger than a critical size, on the other hand, can release more energy than is needed to create their walls.   Those bubbles will expand indefinitely, at an accelerating rate, and ultimately engulf the entire system. But the time needed to create a super-critical bubble can be extremely long: the predicted decay time is, in fact, exponentially long in the inverse of the energy density difference between the true and false vacua \cite{Rutkevich,Voloshin1985,Lagnese_2022,ChaoLucas2023}.
Can we identify observable signatures testing whether a system -- or the universe -- is in a false vacuum even if the decay takes an absurdly long time?

\begin{figure}[t]%
\centering
\includegraphics[width=\columnwidth]{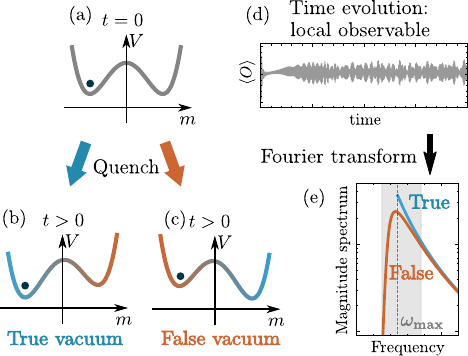}
\caption{(a) The system is prepared in one of the degenerate vacua. The Hamiltonian is quenched such that the state of the system is now close (b) to the true vacuum or  (c) to a false vacuum. 
(d) Real-time evolution of a  local observable
after the quench. (e) In the appropriate range of frequencies (grey region), the Fourier transform to the frequency domain (magnitude spectrum) reveals whether the system was in the true vacuum or a false vacuum. 
}\label{fig1}
\end{figure}

The traditional approach to bubble nucleation and false vacuum decay computes the decay rate using a semiclassical approximation \cite{coleman_1977,Coleman1979}.  In the limit of small quantum fluctuations, the contribution of the dominant, critical bubble can be isolated by solving a classical evolution in imaginary time. Quantum corrections around the semi-classical solution are then included \cite{Callan:1977pt,andreassenPRL,andreassenPRD,Ivanov2022}.
Recently, motivated by advances in the experimental study of out-of-equilibrium dynamics in many-body quantum systems \cite{Hofferberth2007,Trotzky2012,Gring2012,Cheneau2012,Meinert2013,Langen2013,schreiber2015observation,langen2015experimental,Kaufman2016,Fukuhara2013,bernien2017,zhang2017observation,bluvstein2021controlling}, a different method has been used  \cite{Lagnese_2022,Gabor2022,Pomponio2022,Mussardo_2022}.  In this method one focuses on real-time evolution after a sudden change in some parameters of the theory (a quantum quench \cite{calabrese_cardy2006,calabrese_cardy2007}). Quantum quenches are now being studied fruitfully in quantum simulators \cite{bernien2017,SuracePRX,banerjeePRL,vovrosh2021confinement,tan2021domain,zenesini2023observation}, and in (classical) numerical simulations using tensor network techniques \cite{Kormos2017,Lerose2020,Lagnese_ladder,Birnkammer:2022giy,Scopa2022,Milsted2022,maki2023monte}. As exemplified in  Fig.~\ref{fig1}, the quench procedure for studying the decay of a false vacuum begins with a system prepared at time $t=0$ in one of its stable configurations [Fig.~\ref{fig1}-(a)]. A parameter of the Hamiltonian is suddenly changed such that at $t>0$ the system occupies a possible metastable --- false vacuum --- configuration [Fig.~\ref{fig1}-(c)]. This case is contrasted with a similar procedure where the system is in a stable configuration (true vacuum) after the quench, as in Fig~\ref{fig1}-(b).  Starting from one of these new configurations, the dynamics can be characterized by monitoring the time evolution of a suitable observable, as in Fig.~\ref{fig1}-(d) \cite{DELFINO2023116312}.

In this Letter, we propose to investigate the magnitude spectrum of the selected observable [Fig.~\ref{fig1}-(e)]. In the limit of small frequency $\omega \rightarrow 0$, the traditional results for the false vacuum decay rate based on semiclassical approaches are recovered. However, as a consequence of the long lifetime of the false vacuum, the magnitude spectrum has extremely small amplitude in this small $\omega$ regime, hindering the direct detection of the decay.  The main goal of this Letter is to show that the post-quench dynamics can reveal much more information towards the detection of metastability. We attribute this effect to the occurence of subcritical,``virtual'' bubbles: these are off-resonant states in the language of equilibrium perturbation theory, but can be excited in such non-equilibrium setup. These subcritical bubbles provide a sizeable, observable signal in the magnitude spectrum in a finite range of frequencies, as illustrated in Fig.~\ref{fig1}-(e). 
More precisely, we consider the magnitude spectrum close to its maximum, at frequency $\omega_{\text{max}}$ (approaching $\omega_{\text{max}}$ from below): if the magnitude spectrum (either its continuous profile or its discrete sequence of peaks) exhibits a jump below $\omega_{\text{max}}$, we can argue that the state considered is the true vacuum; if, on the other hand, the dependence is smooth, we are in the presence of a false vacuum. 
In the following, we demonstrate those claims quantitatively in a tractable model, where the temporal structure of the post-quench oscillations of the local observables can be computed from an effective model for the dynamics of the bubbles.

\paragraph{Quench protocol ---} We consider the transverse field Ising model with a confining longitudinal field, defined by
\begin{equation}
    H = -\sum_{i} \left[ \sigma_i^z \sigma_{i+1}^z -h_z \sigma^z_i -h_x \sigma^x_i\right].
    \label{eq:ising}
\end{equation}

We initialize the system in the 
product state $\ket{\psi_\downarrow}$ with spins polarized in the $-\hat z$ direction, so that    $ \bra{\psi_\downarrow} \sigma^{z}_i \ket{\psi_\downarrow} = -1 \; \forall i $. Then we prepare the ground state for $h_z=0$ and fixed $h_x$ (with $0<h_x<1$) by evolving $\ket{\psi_\downarrow}$ in imaginary time until we reach convergence. The ground state that we obtain is one of the two degenerate ground state of the quantum Ising chain  with $h_z = 0$ in the ferromagnetic phase, namely the one with negative magnetization, as illustrated in Fig.~\ref{fig1}-(a). Subsequently, this state is evolved in real time with either (i) $h_z>0$, in which case we say we are evolving from the true vacuum $\ket{0_+}$, Fig.~\ref{fig1}-(b);  or (ii) $h_z<0$, evolving from the false vacuum $\ket{0_-}$, Fig.~\ref{fig1}-(c).

We then compute the magnetization during the real-time evolution, i.e., $\braket{S_z(t)}=\sum_i \langle \psi(t)|  \sigma_i^z|\psi (t) \rangle/2N$,
where $\ket{\psi(t)}=e^{-iHt}\ket{0_{\pm}}$ and $N$ is the number of spins in the chain.
The expectation value of the magnetization exhibits oscillations over time, as shown in Fig.~\ref{fig1}-(d). These modes are the main focus of our analysis [Fig.~\ref{fig1}-(e)]. In particular, we will analyze the magnitude spectrum
\begin{equation}
    \left| S_z(\omega) \right| ~=~\left| \frac{1}{T}\int_0^T\mathrm d t\langle S_z (t)\rangle e^{i\omega t} \right|
\end{equation}
The procedure outlined here for a quantum Ising chain can be extended to any model possessing a multiple-folded ground state degeneracy: a suitable small symmetry-breaking field is quenched and the time evolution of the corresponding order parameter is examined~\footnote{Note that an underlying symmetry is not strictly required: the same analysis can be applied near a first order phase transition. In that case, any parameter that can be used to tune across the phase transition plays the role of the symmetry-breaking field.}.

\begin{figure*}[t!]%
\centering
\includegraphics[width=2.0\columnwidth]{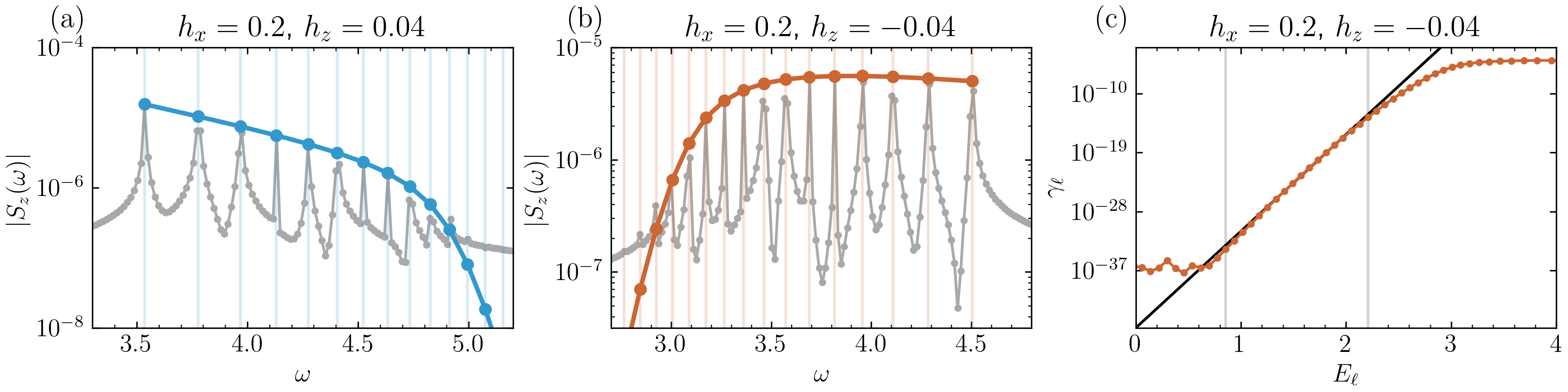}
\caption{\label{fig3} Magnitude spectrum of the real-time evolved magnetization after a quench from (a) the true vacuum and (b) the false vacuum.The grey points are the data obtained from iTEBD, while the blue [for the true vacuum, in (a)] and orange [for the false vacuum, in (b)] points represent  the predicted peaks positions $E_{\ell}$ and amplitudes $|S^z(\omega=\pm E_{\ell})|$ obtained within the two-particle approximation [i.e., extracted from Eq.~(\ref{eq:eigsimproved2}) and (\ref{eq:peakmagn})].  Panel (c): exponential fitting of the amplitudes of the subcritical bubbles and $E_{\ell} \to 0$ extrapolation of the decay rate $\gamma$ according to Eq.~(\ref{eq:rates}). Parameters are $h_x=0.2$ and  $|h_z|=0.04$.}
\end{figure*}

\paragraph{Theory --}
Now we provide a theoretical framework for understanding the post-quench dynamics.  
For $h_z=0$ the Hamiltonian can be mapped exactly onto a free-fermion model. We restrict to $0<h_x<1$, so that the ground state prepared with imaginary time evolution is one of the ferromagnetic ground states \footnote{ When using finite size methods, one must also include a small longitudinal field in order to avoid creating a cat state.}. The dispersion relation of the fermions ({\it kinks}) is 
\begin{equation}
    \epsilon(\theta)=2 \sqrt{(1-h_x)^2 + 4 h_x \sin^2\left(\frac{\theta}{2}\right)},
\end{equation}
where $\theta$ is the quasimomentum of the fermion.


Let us turn on a small $h_z\neq 0$. The longitudinal field $V=\sum_i h_z \sigma_i^z$ induces a linear potential that binds the fermionic excitations in pairs \cite{McCoyWu, Delfino}, and also allows creation and annihilation of pairs of excitations. Following \cite{Rutkevich}, we can split $V$ into two terms. The first, ``force'' term is obtained by projecting $V$ within sectors with a fixed number of fermions. As we will discuss, this term completely changes the spectrum of the model and must be treated non-perturbatively.   The second residual interaction, which couples sectors containing different numbers of fermions, can be treated perturbatively. Since $|h_z|$ is much smaller than the mass $\Delta\equiv 2(1-h_x)$ of the fermions, for our purposes we can restrict our attention to the no-particle and two-particle sector, and also study the two terms separately.

\paragraph{Two-particle spectrum ---}
The energy levels in the two-particle sector are obtained by solving the following eigenvalue equation for a pair of fermions with zero center-of-mass momentum (See the Supplemental Material \cite{supmat} for a derivation):\nocite{landau1991quantum,Fendley2004,Rico2014}

\begin{equation}
\label{eq:eigsimproved2}
    \sum_{n' \in \mathbb{Z}^+} (K_{n-n^{\prime}}-K_{n+n'}) \phi_\ell(n') \pm M |h_z| n \phi_\ell(n)= \frac{E_\ell}{2} \phi_\ell(n).
\end{equation}
Here $n\in \mathbb Z^+$ labels the distance between the fermions, $\phi_\ell (n)$ is the eigenfunction with quantum number $\ell$, $M=(1-h_x^2)^{1/8}$ is the  $h_z=0$ ground state expectation value of the magnetization, and the hoppings $K_r$, derived from the free-fermion Hamiltonian, are defined as
\begin{equation}
K_{r}= \int_{-\pi}^{\pi} \frac{d\theta}{2 \pi} \epsilon(\theta) e^{i r \theta}.
\end{equation}
The quantum number $\ell=1,2,\dots$ labels the eigenstates in ascending order with respect to their characteristic inter-particle distance.
The plus/minus signs {in Eq.~(\ref{eq:eigsimproved2})} are for the quench from the true/false vacuum respectively.
The solutions for the $``+"$ sign can be interpreted as {\it mesons}: they are two-kink bound states that result from the attractive linear potential generated by the longitudinal field; they have positive energy $E_\ell$ with respect to the ground state, with $E_\ell$ increasing with $\ell$, and they represent the new low-energy excitations of the model. The solutions for the $``-"$ sign are more properly named {\it bubbles\/} (though we sometimes use that term in both cases).  They can have positive or negative $E_\ell$, and, since the interaction is repulsive, $E_\ell$ decreases with $\ell$. 
The critical bubble $\ell_\text{res}$ corresponds to the smallest $\ell$ such that $E_\ell\le 0$.

In Fig.~\ref{fig3}-(a) and Fig.~\ref{fig3}-(b),  
we compare the energy levels obtained by solving Eq.~(\ref{eq:eigsimproved2}) with the {magnitude spectrum} $|S_z(\omega)|$ obtained from 
{infinite time-evolving block decimation \cite{Vidal2007,supmat}} (iTEBD)
for the evolution from the true and false vacuum, respectively \review{(see \cite{supmat} for analogous results obtained with different parameters)}. The solutions $E_\ell$ (\review{colored} vertical lines) are in excellent agreement with the frequencies of the modes in $S_z(\omega)$.

\paragraph{Amplitudes ---}
To compute the amplitudes of the oscillations of the magnetization we assume
\begin{equation}
    \ket{\psi(t)} \simeq \ket{0_\pm}+\sum_\ell a_\ell(t) e^{-iE_\ell t} \ket{\phi_\ell},
\end{equation}
where $\ket{0_\pm}$ is the true/false vacuum and $E_\ell$, $\ket{\phi_\ell}$ are the two-particle energies and eigenstates from Eq.~(\ref{eq:eigsimproved2}). The coefficients $a_\ell(t)$ depend on the coupling between the vacuum and the two-particle states induced by the perturbation. By evaluating these matrix elements (see the Supplemental Material \cite{supmat}), we find that
the peaks of $|S_z(\omega)|$ occur at energies $\pm E_\ell$ with amplitude
\begin{equation}
	\lvert S_z(\omega=\pm E_\ell) \rvert \simeq
	 \frac{|h_z| M^2}{8 |E_\ell|} \left(\sum_{n} h_x^n \phi_\ell (n)\right)^2.
    \label{eq:peakmagn}
\end{equation}

In Fig.~\ref{fig3}-(a,b\review{\sout{,d,e}}) these theoretical prediction for the amplitudes (\review{\sout{black} blue/orange} dots) are compared with the amplitudes {in the magnitude spectrum}
of the real-time evolved magnetization \review{(in grey)}.
We find that, within the region where practical limitations on time length and time step allow for clear resolution, the peaks in the spectrum are remarkably consistent. This demonstrates that the dynamics are dominated by the two-particle modes

\paragraph{Decay rate ---} {The decay rate from the false vacuum is dictated by the rate of formation of a resonant bubble. Since the resonant bubble has a size $\ell_{\text{res}}\propto h_z^{-1}$, a simple perturbative estimate of the matrix element for such process (corroborated by more rigorous calculations \cite{Rutkevich}) suggests that the decay rate $\gamma$ is exponentially small in $h_z^{-1}$. We are now going to show that, }
even without a direct observation of the formation of resonant true-vacuum bubbles, it is possible to extrapolate the decay rate of the false vacuum from the peak amplitudes in the magnetization of the non-resonant states. The analytic expression for the decay rate~\cite{Rutkevich,supmat} can be approached using the auxiliary quantities
\begin{equation}
\gamma_\ell\equiv 2\pi\frac{|V_{\ell}|^2}{N}=4 \pi h_z E_{\ell} |S_z(\omega=E_{\ell})|,
    \label{eq:rates}
\end{equation}
where $V_\ell=\bra{0_-} V \ket{\phi_\ell}$. \review{In Fig.~\ref{fig3}-(c) we plot the values of $\gamma_\ell$ (orange dots) computed using the frequencies $E_\ell$ and the amplitudes $|S_z(\omega=E_\ell)|$ obtained from Eqs.~(\ref{eq:eigsimproved2}) and (\ref{eq:peakmagn}).}
The actual decay rate $\gamma$, i.e., the probability density per unit time for the critical bubble to be produced, is obtained from Eq.~(\ref{eq:rates}) in the limit
$E_\ell \to 0$. 
This limit can be extrapolated from a suitable exponential fitting of $\gamma_\ell$\review{\sout{, as illustrated Fig.~\ref{fig3}-(c,f)}}   as a function of the energy level $E_\ell$\review{, as illustrated by the black line in Fig.~\ref{fig3}-(c)}. \review{\sout{(For pragmatic reasons, we inserted the theoretical prediction for $S_z(\omega)$ from Eq.~(\ref{eq:peakmagn}) directly into Eq.~(\ref{eq:rates}).)}}
{The extrapolated exponents of the decay rates are \review{\sout{$\log (\gamma)=-196$ for $h_z=0.02$ [Fig.~\ref{fig3}-(c)] and}} $\log (\gamma)=-108$ for $h_z=0.04$ [Fig.~\ref{fig3}-(\review{\sout{f} c})] \review{$\log (\gamma)=-196$ for $h_z=0.02$ [Fig.~S2-(c) in \cite{supmat}]}. The results are contrasted with the analytical prediction formulated in Ref.~\cite{Rutkevich}, which states the decay exponents are \review{\sout{$\log (\gamma_\mathrm{th}) = -208$  and}} $\log (\gamma_\mathrm{th}) = -104$ \review{and $\log (\gamma_\mathrm{th}) = -208$}. } {Note that such extrapolation shares, by construction, the same underlying assumption behind the computation of the decay rate in Ref.~\cite{Rutkevich} which is known to disagree with other estimates \cite{Voloshin1985,fonseca2003ising} by a constant prefactor (see for instance the discussion in Ref.~\cite{Gabor2022}). Hence, our analysis cannot shed any light on the conflicting results concerning the value of the prefactor. }

Importantly, the characteristic time scale of the decay $\tau\approx 1/\gamma$ is much larger than the evolution time $T=400$ that we use for the iTEBD simulations in Figs.~\ref{fig3}-(a,b\review{\sout{,d,e}}). The {magnitude spectrum analysis} of the oscillations can therefore probe the stable/metastable nature of the vacuum on time scales that are significantly shorter than the decay time (by many orders of magnitude).

\paragraph{Scaling limit --}
We now show how to connect our results for the Fourier transform $S_z(\omega)$ to continuum Ising field theory. The lattice model in Eq.~(\ref{eq:ising}) has a quantum critical point in $h_x=1$, $h_z=0$: in its vicinity, the model scales to the $c=1/2$ Ising conformal field theory, perturbed by the energy density operator $\epsilon(x)$ --- yielding a finite mass $m$ for the kinks --- and the spin operator $\sigma(x)$ as the effect of a longitudinal field $h$ {
\cite{difrancesco,johncardy}}. 
This field theory depends on the single dimensionless parameter $\eta =m\cdot|h|^{-8/15}$. 

 To reach the continuum limit from our results for the quantum Ising chain, we rescale the frequency $\omega$ in units of the mass gap $\Delta=2(1-h_x)$, and the magnetization in units of $M$:
\begin{equation}
    \tilde \omega_\ell \equiv \frac{E_\ell}{\Delta},\hspace{0.4cm}
   \tilde S_\ell\equiv \frac{ S_z(E_\ell)}{M}.
\end{equation}

\begin{figure}[h]
    \centering
    \includegraphics[width=\columnwidth]{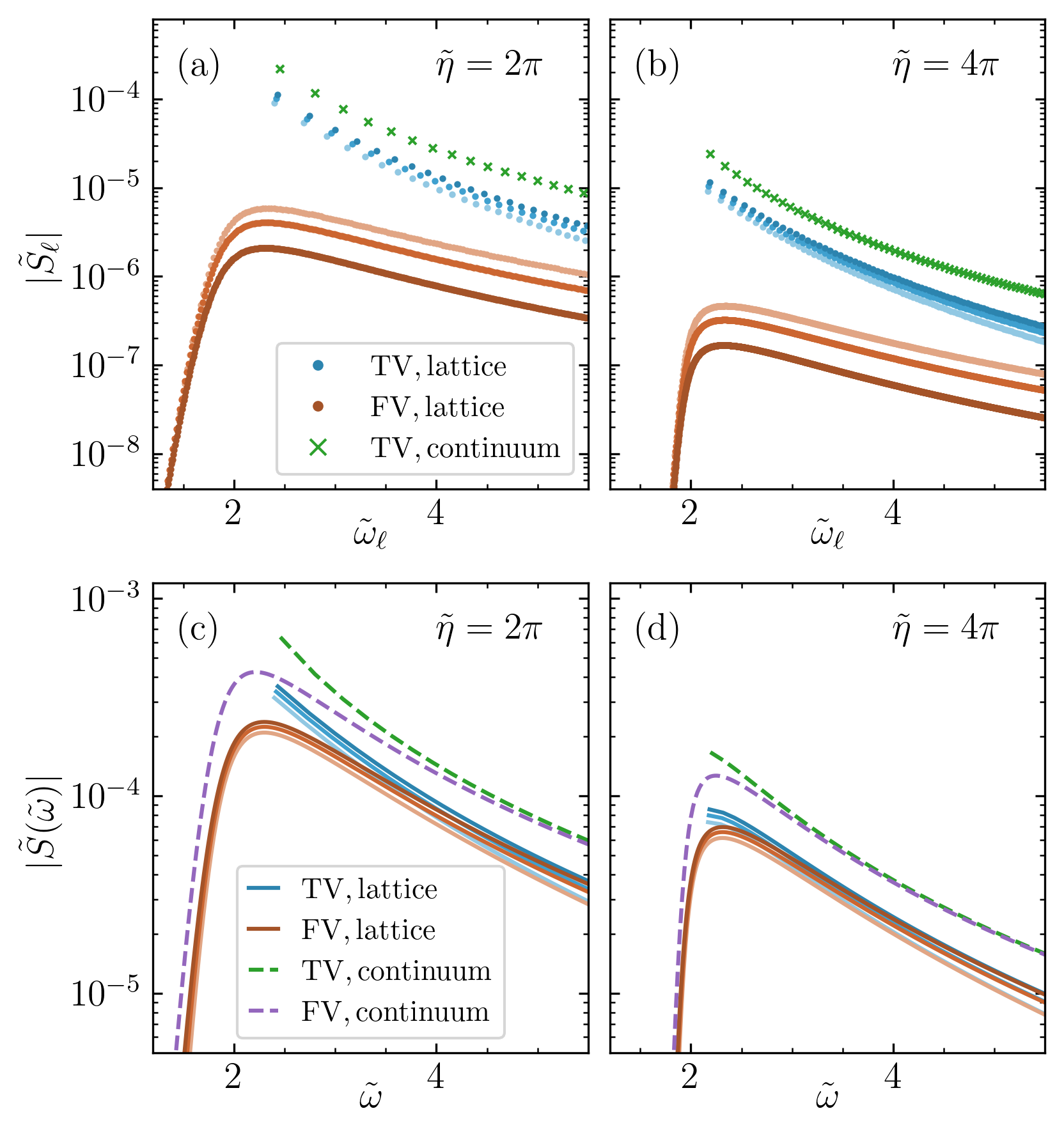}
    \caption{Scaling limit. Blue/Orange: magnetization spectrum for the lattice models. The intensity of the color represents the closeness to the critical point ($h_x=0.7,0.8,0.9$, from lighter to darker tone).  Green/Purple: magnetization spectrum in the continuum for the true/false vacuum within the non-relativistic approximation.}
    \label{fig:scaling limit}
\end{figure}

In Fig.~\ref{fig:scaling limit}-(a,b) we plot our results (obtained using Eq.~(\ref{eq:peakmagn})) for values of $h_x, h_z$ increasingly close to the critical point, while keeping the parameter $\tilde \eta$ defined by
\begin{equation}
    \tilde \eta = (2\bar s)^{8/15}\frac{(1-h_x)}{(1+h_x)^{1/15}h_z^{8/15}},
\end{equation}
fixed.  Here $\bar{s}=1.35783\dots\;$.  The parameter $\tilde \eta$ approximates the dimensionless parameter $\eta$ of the field theory, and $\tilde \eta \rightarrow \eta$ in the scaling limit {\cite{supmat}}.
We observe that the values of $|\tilde S_\ell|$ [Fig.~\ref{fig:scaling limit}-(a,b)] tend to converge to finite values as $h_x\rightarrow 1$ for the case of the true vacuum \review{(blue dots, of increasing darkness for $h_x=0.7, 0.8, 0.9$)}, and are in good agreement with the results obtained in the continuum using the non-relativistic approximation \review{(green crosses)} \cite{supmat}\footnote{Our methods rely on the non-relativistic approximation and the two-fermion Hamiltonian truncation. Note that more accurate predictions for the post-quench dynamics in the continuum can be obtained using efficient numerical methods such as the Truncated Fermionic Space Approach (TFSA) \cite{RAKOVSZKY2016805}.}. For the false vacuum, on the other hand, the values of $|\tilde S_\ell|$ tend to zero for $h_x\rightarrow 1$ \review{(orange dots, of increasing darkness)}. The reason is that, while the spectrum of the mesons remains discrete in the continuous limit, the spectrum of the proper bubbles becomes continuous: the typical gap in the (rescaled) spectrum of the bubbles is $|\tilde \omega_\ell-\tilde \omega_{\ell-1}| \sim 2h_zM/\Delta=2\bar s \tilde \eta^{-15/8}(1-h_x)$, that goes to zero in the scaling limit.
We can obtain the continuous spectrum of the magnetization by an appropriate rescaling bringing in the density of states $\rho(\omega_\ell)\sim (\tilde \omega_{\ell+1}-\tilde \omega_\ell)^{-1}$:
\begin{equation}
    \tilde \omega_\ell\rightarrow \tilde\omega,\hspace{0.4cm} \tilde S_\ell\cdot (\tilde \omega_{\ell+1}-\tilde \omega_\ell)^{-1} \rightarrow \tilde S(\tilde \omega).
\end{equation}
The plots in Fig.~\ref{fig:scaling limit}-(c,d) show the convergence of $\tilde S(\tilde \omega)$ for \review{both the true vacuum (blue lines)} the false vacuum \review{(orange lines). Moreover, the profiles are in good agreement with the ones obtained in the continuum in the non-relativistic approximation \cite{supmat} (green/purple dashed lines for the true/false vacuum respectively)}. 

Observe that the profiles of $\tilde S(\tilde \omega)$ for the two cases (true and false vacuum) tend to coincide for large $\tilde \eta$ in the range $\tilde \omega\gtrsim 2$. We can understand this as follows. The limit $\eta\rightarrow \infty$ corresponds to the quench with a vanishingly small longitudinal field $h$ (known as {\it thin wall} limit). In this limit, we recover the standard theory of false vacuum decay: metastability has observable effects only at long times (longer than the inverse of the mass of two kinks), while the short-time dynamics is governed by the free dynamics of the kinks, which are not affected by the very weak confining potential at short times \footnote{Note that the mesonic spectrum becomes continuous in the limit $\eta\rightarrow\infty$ \cite{fonseca2006ising}.}.\nocite{fonseca2006ising} On the other hand, if $\eta$ is finite, the dynamics of the magnetization can reveal whether the initial state was the true or the false vacuum.
We should remark, however, that our truncation at two fermions precludes accurate treatment of very small $\eta$.  

{Fig.~\ref{fig:scaling limit}-(c,d) shows that, while the spectra agree at large frequencies, and they both vanish at small frequencies, they differ for frequencies comparable to twice the particle mass. The true vacuum has a discontinuous spectrum, which is understood from the presence of a finite energy gap. On the other hand, the magnitude spectrum for the false vacuum shows a smooth dependence on the frequency. In fact, it shows a fast decay for decreasing frequencies in the range $\tilde \omega <2$ but with no sudden jumps. Notably, the difference is observable close to the largest amplitude of the magnitude spectrum, where the signal to noise ratio is most favorable. }

\review{\sout{This suggests a general practical procedure based on investigating the magnitude spectrum close to its maximum, at frequency $\omega_{\text{max}}$ (approaching $\omega_{\text{max}}$ from below): if the magnitude spectrum (either its continuous profile or its discrete sequence of peaks) exhibits a jump below $\omega_{\text{max}}$, we can argue that the state considered is the true vacuum; if, on the other hand, the decay is smooth, we are in the presence of a false vacuum.}}

\paragraph{Outlook --}

The  analysis performed here for the paradigmatic example of the quantum Ising chain can be applied to probe the presence of a long-lived false vacuum in a very general class of models. In \cite{supmat}, we show how this method allows to detect the signatures of a long-lived false vacuum in a Rydberg atom array on time scales that are experimentally accessible. The issue of discriminating between stability and long-lived metastability arises very generally as one varies parameters in systems that exhibit first-order phase transitions, particularly near
the onset and terminal structures of hysteresis curves. We foresee immediate applications of this procedure to other experimental platforms, including ultracold atoms in optical lattices, trapped ions, and superconducting qubits {\cite{GrossBloch2017,bernien2017,browaeys2020many,MonroeReview,tan2021domain,morvan2022formation,zenesini2023observation}, where quench experiments are routinely performed}.  

A phenomenon 
closely related to the decay of the false vacuum was observed in the string breaking dynamics after a quench in quantum spin chains~\cite{Lerose2020,verdel2023dynamical}, where the unstable string connecting two charges can persist for  a very long time. We expect that a similar {magnitude spectrum analysis} could help interpret the real-time evolution observed in these works.

Promising directions include, moreover, the study of metastable phases of matter and their dynamical preparation. A metastable quantum spin liquid phase could explain, for example, the features recently observed in a Rydberg-atom quantum simulator in two spatial dimensions \cite{semeghini2021probing,Giudici2022,sahay2022quantum}.

In the one-dimensional system we focused on here, the boundary between true and false vacuum is marked by zero-dimensional structures, i.e., particle-like ``domain walls''.   In higher dimensions the boundaries will be extended objects, having significant low-energy internal degrees of freedom,  notably including  variables that encode their shapes.   In two and three dimensions the boundaries of  near-critical bubbles will be long-lived strings and membranes respectively. Although directly accessing these objects through simple quenching can be challenging, studying smaller and more accessible bubbles should provide valuable insights into their properties.

Also worthy of exploration is the possibility of more structured quenches.   For definiteness, let us consider again an Ising ferromagnet, starting with the ground state having global magnetization downward, where our destabilizing quench consists of turning on a small field favoring upward magnetization.   If at the same time we also flip the spins within a geometrically defined region (or several such), we can encourage the production of large bubbles.  We could also flip only a fraction of the spins, to match the desired magnetization, or other refinements.  Systematic study and use of this ``bubble seeding'' concept is evidently not restricted to the Ising ferromagnet, and opens a wide field for investigation.

Expanding bubble walls implement time-dependent boundary conditions for external fields on an accelerating surface.  Thus they implement the sort of ``moving mirror'' boundary conditions that have been widely proposed as idealizations of black hole horizons \cite{1977RSPSA.356..237D,PhysRevD.31.767}, specifically in connection with Hawking radiation \cite{PhysRevD.91.124060,PhysRevD.36.2327,wilczek1993quantum,PhysRevD.101.025012}. Collapsing bubbles are the black hole analogs. Here, since the analog horizon moves outward, we come upon a wide class of accessible {\it white\/} holes.   This concept can be taken much further, as we shall report elsewhere.

 Finally let us remark that in a cosmological context differences between scalar-induced energy densities (``cosmological constant'' or ``dark energy'') introduces a new feature: regions with higher energy density expand faster.   It has long been known that this phenomenon can stabilize otherwise unstable vacua \cite{coleman1980}, roughly speaking by enabling more rapid expansion to outpace bubble expansion; indeed, this presented a serious difficulty for early models of inflation \cite{guth1981}.  It will be interesting to explore consequences of this new feature for the signatures dynamics discussed here.

\begin{acknowledgments}
\paragraph{Acknowledgments --} 
We acknowledge useful discussions with A. Bastianello, P. Calabrese, S. B. Rutkevich, and G. Takács.

FMS acknowledges support provided by the U.S.\ Department of Energy Office of Science, Office of Advanced Scientific Computing Research, (DE-SC0020290), by Amazon Web Services, AWS Quantum Program, and by the DOE QuantISED program through the theory  consortium ``Intersections of QIS and Theoretical Particle Physics'' at Fermilab.  FW is supported by the U.S. Department of Energy under grant Contract  Number DE-SC0012567, by the European 
Research Council under grant 742104, and by the Swedish Research Council under Contract No. 335-2014-7424.
G.L. would like to express his gratitude to Niccol\'o Maffezzoli from Ca' Foscari University of Venice for his support and continuous encouragement. G.L. acknowledges the support by the P1-0044 program of the Slovenian Research Agency, the QuantERA grants QuSiED and T-NiSQ by MVZI, QuantERA II JTC 2021, and ERC StG 2022 project DrumS, Grant Agreement No. 101077265.

\end{acknowledgments}

\bibliography{bib}
\end{document}


\title{Supplemental material\\ Detecting a long lived false vacuum with quantum quenches}

\author{Gianluca Lagnese}
\affiliation{Institute of Polar Sciences CNR, Via Torino 155, 30172 Mestre-Venezia, Italy}
\affiliation{SISSA, via Bonomea 265, 34136 Trieste, Italy}
\author{Federica Maria Surace}
\affiliation{Department of Physics and Institute for Quantum Information and Matter,
California Institute of Technology, Pasadena, California 91125, USA}
\author{Sid Morampudi}
\affiliation{Center for Theoretical Physics, Massachusetts
Institute of Technology, Cambridge, MA 02139, USA}
\author{Frank Wilczek}
\affiliation{Center for Theoretical Physics, Massachusetts
Institute of Technology, Cambridge, MA 02139, USA}
\affiliation{T. D. Lee Institute and Wilczek Quantum Center,
SJTU, Shanghai}
\affiliation{Arizona State University,
Tempe AZ USA}
\affiliation{Stockholm University
Stockholm, Sweden}

\date{\today}

\maketitle

\setcounter{equation}{0}
\setcounter{figure}{0}
\setcounter{table}{0}
\setcounter{page}{1}
\makeatletter
\renewcommand{\theequation}{S\arabic{equation}}
\renewcommand{\thefigure}{S\arabic{figure}}
\renewcommand{\bibnumfmt}[1]{[S#1]}
\renewcommand{\citenumfont}[1]{S#1}

In this Supplemental Material, we discuss in detail the derivation of the results shown in the main text. In Section \ref{secA1} we introduce the two-particle model used to compute the spectrum of mesons/bubbles in the quantum Ising chain in a transverse and a (small) longitudinal field; we then use the solutions to derive the Fourier spectrum of the magnetization after the quench [Eq.~(7) in the main text]. In Section \ref{sec:IFT} we introduce the Ising field theory, which describes the quantum Ising chain in the proximity of the critical point; we solve the two-particle model in the continuum in the non-relativistic approximation, and we then show how to use the solutions to compute the Fourier spectrum of the magnetization for the quantum Ising chain in the scaling limit. In Section \ref{sec:Rydberg} we show how to use a similar protocol (as the one proposed in the main text for the quantum Ising chain) to probe the vacuum stability in a Rydberg atom chain.

\section{Two particle spectrum and oscillation amplitudes}\label{secA1}

In the derivation of the two-particle model, we follow Ref.~\onlinecite{Rutkevich}, and define the wavefunction $\phi_\ell(n)$ where $\ell$ labels the eigenstates and $n$ is the distance in real space of the two fermions (with center of mass momentum $k_\mathrm{tot}=0$).
The energy levels are obtained solving the equation
\begin{equation}
\label{eq:eigsimproved}
    \sum_{n' \in \mathbb{Z}} K_{n-n^{\prime}} \phi_\ell(n') \pm M h_z n \phi_\ell(n)= \frac{E_\ell}{2} \phi_\ell(n), 
\end{equation}
where 
\begin{equation}
K_{r}= \int_{-\pi}^{\pi} \frac{d\theta}{2 \pi} \epsilon(\theta) e^{i r \theta},
\qquad
    \epsilon(\theta)=2 \sqrt{(1-h_x)^2 + 4 h_x \sin^2\left(\frac{\theta}{2}\right)}.
\end{equation}
$\epsilon(\theta)$ is the dispersion relation of the fermions in the transverse field Ising model. The term $K_{n-n'}$ is obtained by writing the transverse field Ising model as a diagonal free fermion Hamiltonian in momentum space, and Fourier transforming to real space. The term proportional to $h_z$ is the linear potential between the fermions generated by the longitudinal field. $M=(1-h_x^2)^{1/8}$ is the magnetization, and the plus/minus signs are respectively for two fermions in the true vacuum (attractive potential) or in the false vacuum (repulsive potential).
The two-domain-wall wavefunctions are antisymmetric, with $\phi_\ell(n)=-\phi_\ell(-n)$, so we can equivalently solve the following equation for $n>0$

\begin{equation}
\label{eq:eigsimproved2}
    \sum_{n' \in \mathbb{Z}^+} (K_{n-n^{\prime}}-K_{n+n'}) \phi_\ell(n') \pm M h_z n \phi_\ell(n)= \frac{E_\ell}{2} \phi_\ell(n).
\end{equation}

This equation can be efficiently solved numerically by introducing a cutoff $0<n, n'<N_{\max}$, and checking the convergence in $N_{\max}$.
From the knowledge of the eigenfunctions $\phi_\ell(n)$ it is possible to extract the matrix elements of the long range interaction term from (see \onlinecite{Rutkevich}) 
\begin{equation}
    \bra{0_{\pm }}V\ket{\phi_\ell} = i h_z M \sqrt{N} \int_{-\pi}^{\pi} \frac{d \theta}{2 \pi} \phi_\ell(\theta) \frac{d \log \epsilon (\theta)}{d \theta},
\end{equation}
where $V=h_z\sum_i \sigma_i^z $ and $N$ is the system size.
Using that $\phi_\ell(\theta) = \sum_n e^{i n \theta} \phi_\ell(n)$ and performing the integration, the result reads
\begin{equation}
\label{eq:matrixel}
    \bra{ 0_{\pm }}V\ket{\phi_l} =  \frac{1}{2} M \sqrt{N} h_z \sum_n h_x^n \phi_\ell(n).
\end{equation}

We can now estimate the time evolution of the magnetization as
\begin{equation}\label{eq:magnetization}
    \frac{1}{N}\sum_i\langle S_i^z (t) \rangle=\frac{1}{2h_z N}\langle \psi(t)| V|\psi (t) \rangle
\end{equation}
where $\ket{\psi(t)}=e^{-iHt}\ket{0_{\pm}}$.

To lowest order in perturbation theory we can assume
\begin{equation}
    \ket{\psi(t)} \simeq \ket{0_\pm}+\sum_\ell a_\ell(t) e^{-iE_\ell t} \ket{\phi_\ell}
\end{equation}
where $a_\ell(t)\simeq \bra{ \phi_\ell} V \ket{0\pm}(e^{iE_\ell t}-1)/(iE_\ell)$.

Keeping only the first order, we obtain

\begin{align}
    \bra{\psi(t)} V \ket{\psi(t)} &\simeq \bra{ 0_\pm } V \ket{0_\pm } +\sum_\ell \left[ a_\ell(t) e^{-iE_\ell t} \bra{0_\pm} V \ket{\phi_\ell}+\text{c.c.} \right]\\
    &\simeq \bra{0_\pm } V \ket{ 0_\pm} +\sum_\ell \left[ \frac{1-e^{-iE_\ell t}}{iE_\ell}  
    \lvert \bra{0_\pm}V\ket{\phi_\ell}\rvert^2+\text{c.c.} 
    \right]
    \\
    &= \bra{0_\pm } V \ket{0_\pm } +\sum_\ell  \frac{2\sin E_\ell t}{E_\ell}  \lvert\bra{0_\pm} V \ket{\phi_\ell}\rvert^2.
\end{align}

From this equation, we obtain that the Fourier transform $S^z(\omega)=\frac{1}{TN}\int_0^T\mathrm d t \sum_i\langle S^z_i (t)\rangle e^{i\omega t} $ for infinite $T$ has peaks at energies $\pm E_\ell$ of amplitude
\begin{equation}
\lvert S^z(\omega=E_\ell) \rvert \simeq\frac{1}{2 h_z E_\ell} \frac{\lvert \bra{0_{\pm }}V \ket{\phi_\ell}\rvert^2}{N}=\frac{h_z M^2}{8 E_\ell} \left(\sum_{n} h_x^n \phi_\ell (n)\right)^2.
    \label{eq:peakmagn}
\end{equation}

\section{Ising field theory and scaling limit}
\label{sec:IFT}
The scaling limit of the quantum Ising chain in transverse and longitudinal field is described by the Ising Field Theory (IFT) \cite{Delfino,fonseca2003ising,fonseca2006ising}: this corresponds to the Ising $c=1/2$ conformal field theory perturbed by two relevant operators $\varepsilon(x)$ (energy density, with conformal dimension $1/2$) and $\sigma(x)$ (spin density, with conformal dimensions $1/16$). The Euclidean action in two dimensions reads
\begin{equation}
    \mathcal A_{\mathrm{IFT}}=\mathcal A_{c=1/2,\; \mathrm{CFT}}+\tau \int \mathrm{d}^2 x\, \varepsilon(x)+h \int \mathrm{d}^2 x\, \sigma(x).
\end{equation}
The theory depends on a single adimensional parameter, defined as

\begin{equation}
    \eta= \frac{2\pi \tau}{h^{8/15}}= \frac{m}{h^{8/15}}.
\end{equation}

\subsection{Two-particle model in the continuum}
\label{sec:nonrel}
We are interested in comparing the results obtained for the quantum Ising chain using the two-fermion approximation with the analogous calculation in the IFT. We therefore focus on the following
two-body problem which effectively describes the Ising scaling limit \cite{fonseca2006ising}

\begin{equation}
    H=\omega(p_1)+\omega(p_2)+2h\bar \sigma |x_1-x_2|
\end{equation}
where $\omega(p)=\sqrt{p^2+m^2}$ can be approximated in the non-relativistic limit with $\omega(p)\simeq m+ \frac{p^2}{2m}$. The non-relativistic approximation is valid for large $\eta$, which is consistent with our regime of interest. The linear potential is attractive ($h>0$) for pairs of particles on top of the true vacuum, and repulsive ($h<0$) for pairs of particles on top of the false vacuum. Changing variables to the center of mass and relative coordinates we get 

\begin{equation}
    H=2m+\frac{P^2}{4m}+\frac{p^2}{m}+2\bar \sigma h |x|,
\end{equation}
where
\begin{align}
    &X=\frac{x_1+x_2}{2}, \qquad &P=p_1+p_2,\\
    &x=x_1-x_2, \qquad &p=\frac{p_1-p_2}{2}.
\end{align}

The eigevalues of $H$ are given by $E^{\mathrm{tot}}_{P,\ell}=2m+\frac{P^2}{4m}+E_\ell$ where $E_\ell$ are obtained solving the eigenvalue problem for the relative coordinate:
\begin{equation}
    \left(-\frac{1}{m}\partial_x^2+2\bar \sigma h x\right)\psi_\ell(x)=E_\ell \psi_\ell(x),
\end{equation}
with the boundary condition $\psi_\ell(0)=0$. {The latter equation represents the direct result of the scaling limit of Eq.~\ref{eq:eigsimproved}, once the non-relativistic approximation is applied.} {With the} change of variable $z=x/c-c^2mE_\ell$, where $c = (2m\bar \sigma h)^{-1/3}$, {we get}
\begin{equation}
\label{eq:airyeq}
    -\partial_z^2 \psi_\ell(cz+c^3mE_\ell)+z\psi_\ell(cz+c^3mE_\ell)=0.
\end{equation}
The solutions of Eq. (\ref{eq:airyeq}) can be expressed in terms of the Airy functions of first and second kind:
\begin{equation}
    \psi_\ell (x)=a_\ell \cdot \mathrm{Ai}\left(\frac{x}{c}-c^2mE_\ell\right)+b_\ell \cdot \mathrm{Bi}\left(\frac{x}{c}-c^2mE_\ell\right).
\end{equation}
 The coefficient $a_\ell$, $b_\ell$ have to be fixed such that the boundary condition $\psi_\ell(0)=0$ is satisfied, and that $\psi_\ell(x)$ does not diverge at large (positive) $x$. As we show below, the solutions depend on the sign of $h$, yielding a discrete spectrum (mesons) for $h>0$, and a continuous spectrum (bubbles) for $h<0$.

\subsubsection{Mesons ($h>0$)}
For the case of excitations on top of the true vacuum, we have $h>0$ and thus $c>0$. Since $\mathrm{Bi}(x)$ diverge at $x\rightarrow +\infty$ we set $b_\ell=0$. The condition $\psi_\ell(0)=0$ gives the condition on the eigenvalues $E_\ell$:
\begin{equation}
    \mathrm{Ai}\left(-c^2mE_\ell\right)=0\; \Longrightarrow\; E_\ell= -\frac{(2\bar \sigma h)^{2/3}}{m^{1/3}}z_\ell.
\end{equation}
The coefficient $a_\ell$ can be chosen to satisfy the normalization condition

\begin{equation}
    1=\int_0^\infty \mathrm{d}x \left[\psi_\ell(x)\right]^2 = a_\ell^2\int_{0}^\infty \mathrm{d}x \left[\mathrm{Ai}\left(\frac{x}{c}+z_\ell\right)\right]^2=a_\ell^2 c \int_{z_\ell}^\infty \mathrm{d}y \left[\mathrm{Ai}\left(y\right)\right]^2=a_\ell^2 c{\mathrm{Ai}'}^2(z_\ell).
\end{equation}
{The last equality is obtained integrating by part and using the Airy equation. The final result reads}
\begin{equation}
    a_\ell = [\mathrm{Ai}'(z_\ell)]^{-1}c^{-1/2},\qquad \psi_\ell(x)=c^{-1/2}\frac{\mathrm{Ai}\left(\frac{x}{c} +z_\ell\right)}{\mathrm{Ai}'(z_\ell)}.
\end{equation}

\subsubsection{False vacuum bubbles ($h<0$)}
For pairs of excitations produced on the false vacuum, we have $h<0$ and $c<0$: since $\mathrm{Ai}(x)$, $\mathrm{Bi}(x)$ do not diverge for $x\rightarrow -\infty$, both $a_\ell$ and $b_\ell$ can be non-zero. There is a continuum of solutions, that we now label by their energy $E$
\begin{equation}
\label{eq:cont_bubble}
    \psi_E(x)=\mathcal N(E) \cos \theta(E) \cdot \mathrm{Ai}\left(\frac{x}{c}-c^2mE\right)-\mathcal N(E) \sin \theta(E) \cdot \mathrm{Bi}\left(\frac{x}{c}-c^2mE\right)
\end{equation}
where the condition $\psi_E(0)=0$ is used to fix $\theta(E)$: 
\begin{equation}
    \tan \theta(E)= \frac{\mathrm{Ai}(-c^2mE)}{\mathrm{Bi}(-c^2mE)}.
\end{equation}

We can then fix $\mathcal N(E)$ using the normalization condition
\begin{equation}
    \delta(E-E')=\int_0^\infty \mathrm{d}x \, \psi_E(x)\psi_{E'}(x).
\end{equation}

Computing this integral is not straightforward, so we use the method in Ref. \onlinecite{landau1991quantum}: we write the asymptotic solution at large $x$ as the superposition of an incoming and an outgoing wave, and we choose $\mathcal N (E)$ such that the probability current density of the one travelling away from (towards) the origin is $1/2\pi$ ($-1/2\pi$).

The asymptotic expansion of the Airy functions at large $z$ yields:
\begin{equation}
    \mathrm{Ai}(-z)\sim \frac{\sin \left(\frac{2}{3} z^{3/2}+\frac{\pi}{4}\right)}{\sqrt{\pi}|z|^{1/4}}, \qquad \mathrm{Bi}(-z)\sim \frac{\cos \left(\frac{2}{3} z^{3/2}+\frac{\pi}{4}\right)}{\sqrt{\pi}|z|^{1/4}}.
\end{equation}

For large $x$, the wavefunction $\psi_E(x)$ has the asymptotic form
\begin{equation}
    \psi_E(x)\sim\psi_E^{(+)}(x)+\psi_E^{(-)}(x),
\end{equation}
where
\begin{equation}
    \psi_E^{(\pm)}(x)= \mathcal N(E)\frac{|c|^{1/4}}{2\sqrt{\pi}x^{1/4}}(\mp i\cos\theta-\sin\theta)\exp\left[\pm i\left(\frac{2x^{3/2}}{3|c|^{3/2}}+\frac{\pi}{4}\right)\right].
\end{equation}
We use the following expression
\begin{equation}
    \partial_x \psi_E^{(\pm)}(x) =\left(-\frac{1}{4x}\pm i \frac{x^{1/2}}{|c|^{3/2}}\right)\psi_E^{(\pm)}(x)
\end{equation}
and compute the probability current of each component as
\begin{align}
    j^{(\pm)}&= -\frac{i}{m}\left(\psi_E^{(\pm)*}\partial_x \psi_E^{(\pm)}-\psi_E^{(\pm)*}\partial_x \psi_E^{(\pm)}\right)\\
    &=-\frac{i}{m}\left(\pm 2i\frac{x^{1/2}}{|c|^{3/2}}\right)|\psi_E^{\pm}|^2\\
    &=\pm\frac{2}{m}\frac{x^{1/2}}{|c|^{3/2}}\mathcal [\mathcal N(E)]^2\frac{|c|^{1/2}}{4\pi x^{1/2}}\\
    &=\pm \frac{1}{2\pi |c| m}[\mathcal N(E)]^2.
\end{align}
Imposing $j^{(\pm)}=\pm 1/2\pi$, we get $\mathcal N(E) =\sqrt{|c|m}$.

\subsection{Scaling Limit of the quantum Ising chain}
We now want to compare our results for the quantum Ising chain to the Ising field theory in the non-relativistic limit.
To perform this comparison, we need to study the appropriate scaling of the parameters of the quantum Ising chain.
In order to proceed for the scaling limit it is useful to reintroduce the energy scale $J$ in the Hamiltonian of the quantum Ising chain:

\begin{equation}
    H = -J\sum_{i} \left[ \sigma_i^z \sigma_{i+1}^z -h_z \sigma^z_i -h_x \sigma^x_i\right].
    \label{eq:ising}
\end{equation}
The scale $J$ can be expressed in terms of the lattice spacing $a$ as $J=1/2a$. In the scaling limit, the transverse field is sent to $1$, while keeping the mass gap constant, i.e.
\begin{equation}
    J\rightarrow \infty, \qquad h_x\rightarrow 1,\qquad 2J|1-h_x|\rightarrow m =\text{const.}
\end{equation}

We now want to derive the scaling limit of the expression for the Fourier transform of the magnetization found in Sec.~\ref{secA1}.
The expression should depend on the adimensional parameter of the field theory
\begin{equation}
    \eta = \frac{m}{h^{8/15}}.
\end{equation}
It is useful to define also
\begin{equation}
    \lambda=\frac{2\bar{\sigma}h}{m^2}=\frac{2\bar{s}m^{1/8}h}{m^2}=\frac{2\bar{s}h}{m^{15/8}}=2\bar{s}\eta^{-15/8}
\end{equation}
where we used the relation between the magnetization and the mass gap $\bar{\sigma}=\bar{s}m^{1/8}$ with $\bar{s}=1.35783\dots$ \cite{fonseca2003ising}.
Our goal is to compute the continuum limit of the quantity
\begin{equation}
\label{eq:Stilde}
    \tilde S_\ell = \frac{S_z(E_\ell)}{M}=\frac{|h_z| M J }{(1-h_x)8J\tilde \omega_\ell}\frac{\left[\sum_n h_x^n \phi_\ell(n)\right]^2}{\sum_n  \left[\phi_\ell(n)\right]^2},
\end{equation}
where $E_\ell =m \tilde \omega_\ell = 2J(1-h_x)\tilde \omega_\ell $ and we introduced explicitly the normalization of the wavefunctions.
We first note that in the scaling limit
\begin{equation}
\label{eq:sl3}
    \frac{h_zM}{(1-h_x)^2}\rightarrow \lambda.
\end{equation}
This can be shown from the following identifications:
\begin{equation}
\label{eq:sl1}
    JNh_zM\rightarrow hL\bar{\sigma},
\end{equation}
\begin{equation}
\label{eq:sl2}
    2J(1-h_x)\rightarrow m.
\end{equation}
Taking the ratio between Eq.~(\ref{eq:sl1}) and the square of Eq.~(\ref{eq:sl2}), and using $N=L/a=2JL$ we get
\begin{equation}
    \frac{JNh_zM}{4J^2(1-h_x)^2}=L\frac{h_z M}{2(1-h_x)^2}\rightarrow L\frac{h\bar{\sigma}}{m^2}=\frac{1}{2}L\lambda,
\end{equation}
which is equivalent to Eq.~(\ref{eq:sl3}).
Moreover, we can take the continuum limit of the sums in Eq.~(\ref{eq:Stilde}) by replacing them with integrals
\begin{equation}
\label{eq:s1}
    \sum_n h_x^n \phi_\ell(n) \rightarrow a^{-1}\int_0^{+\infty} \mathrm d x\; (1-ma)^{-mx/a}\phi_\ell (x/a)\rightarrow a^{-1}\int_0^{+\infty} \mathrm d x\; e^{-mx}\phi_\ell (x/a),
\end{equation}
\begin{equation}
\label{eq:s2}
    \sum_n  \left[\phi_\ell(n) \right]^2 \rightarrow a^{-1}\int_0^{+\infty} \mathrm d x\; \left[\phi_\ell (x/a)\right]^2,
\end{equation}
with the identification $x=n a$.

The wavefunctions of the mesons can be obtained in the field theory in the non-relativistic limit as shown in Sec. \ref{sec:nonrel}:
\begin{equation}
\label{eq:phil}
    \phi_\ell(x/a)\rightarrow \psi_\ell(x) =\mathcal N^{-1/2} \mathrm{Ai}\left[(2m\bar \sigma h)^{1/3}x+z_\ell\right]=\mathcal N^{-1/2} \mathrm{Ai}\left(\lambda^{1/3}m x+z_\ell\right),
\end{equation}
\begin{equation}
\label{eq:oml}
    \tilde \omega_\ell = \frac{2m+E_\ell}{m} =2-\frac{(2\bar \sigma h)^{2/3}}{m^{4/3}}z_\ell=2-\lambda^{2/3}z_\ell
\end{equation}
where $z_\ell$ is the $\ell$-th zero of the Airy function and $\mathcal N$ is a normalization constant.

Finally, plugging Eqs.~(\ref{eq:sl3}), (\ref{eq:s1}), (\ref{eq:s2}), (\ref{eq:phil}) and (\ref{eq:oml}) in Eq.~(\ref{eq:Stilde}) we get the final result

\begin{align}
    \tilde S_\ell &\rightarrow \frac{\lambda(1-h_x)}{8(2-\lambda^{2/3}z_\ell)}a^{-1}\frac{\left[\int_0^{+\infty} \mathrm {d} x\, e^{-mx} \mathrm{Ai}\left(\lambda^{1/3}m x+z_\ell\right)\right]^2}{  \int_0^{+\infty} \mathrm {d} x\,  \mathrm{Ai}\left(\lambda^{1/3}m x+z_\ell\right)^2}\\
    &=\frac{\lambda}{8(2-\lambda^{2/3}z_\ell)}(1-h_x)(\lambda^{1/3}ma)^{-1}\frac{\left[\int_0^{+\infty} \mathrm {d} y\,  \exp\left(-\lambda^{-1/3} y\right) \mathrm{Ai}\left(y+z_\ell\right)\right]^2}{  \int_0^{+\infty} \mathrm {d} y\,  \mathrm{Ai}\left(y+z_\ell\right)^2}\\
    &\rightarrow \frac{1}{8(2\lambda^{-2/3}-z_\ell)}\frac{\left[\int_0^{+\infty} \mathrm {d} y\, \exp\left(-\lambda^{-1/3} y\right) \mathrm{Ai}\left(y+z_\ell\right)\right]^2}{  \mathrm{Ai}'^2(z_\ell)}.
\end{align}
where $\mathrm{Ai^{\prime}}$ stands for the first derivative of $\mathrm{Ai}$.
To simplify the denominator, we used the following property of the Airy functions:
\begin{align}
    \int_0^{\infty} \mathrm{d}y \, \mathrm{Ai}^2(y+z_\ell) &=  \int_{z_\ell}^{\infty} \mathrm{d}s \, \mathrm{Ai}^2(s)\\
    &= s\mathrm{Ai}^2(s)\Big|_{z_\ell}^\infty - 2\int_{z_\ell}^{\infty} \mathrm{d}s \, s\,\mathrm{Ai}(s)\mathrm{Ai}^\prime(s)\\
    &=-z_\ell \mathrm{Ai}^2(z_\ell)- 2\int_{z_\ell}^{\infty} \mathrm{d}s \, \mathrm{Ai}^{\prime\prime}(s)\mathrm{Ai}^{\prime}(s)\\
    &=-z_\ell \mathrm{Ai}^2(z_\ell) - \mathrm{Ai}^{\prime \, 2}(s)\Big|_{z_\ell}^\infty=\mathrm{Ai}^{\prime \, 2}(z_\ell),
\end{align}
where the last equality follows from the definition of $z_\ell$.

For the case of the continuous spectrum of false vacuum bubbles we can perform a similar calculation for the spectral density $\tilde S(\tilde \omega)$. Using Eq.~(\ref{eq:cont_bubble}), we get

\begin{equation}
    \tilde S(\tilde \omega)=\frac{1}{8\tilde \omega}\left[\cos(\theta(\tilde \omega))\int_0^\infty \mathrm{d}y\, \exp\left(-\lambda^{-1/3}y\right)\mathrm{Ai}(-y+z(\tilde\omega))-\sin(\theta(\tilde \omega))\int_0^\infty \mathrm{d}y\,\exp\left(-\lambda^{-1/3}y\right)\mathrm{Bi}(-y+z(\tilde\omega))\right]^2
\end{equation}
where
\begin{equation}
    z(\tilde\omega) = -\lambda^{-2/3}(\tilde \omega-2), \qquad \theta(\tilde \omega)=\arctan\left(\frac{\mathrm{Ai}(z(\tilde\omega))}{\mathrm{Bi}(z(\tilde\omega)}\right).
\end{equation}

\section{Details on the numerical method}
\begin{figure}
    \centering
    \includegraphics[width =0.9\textwidth]{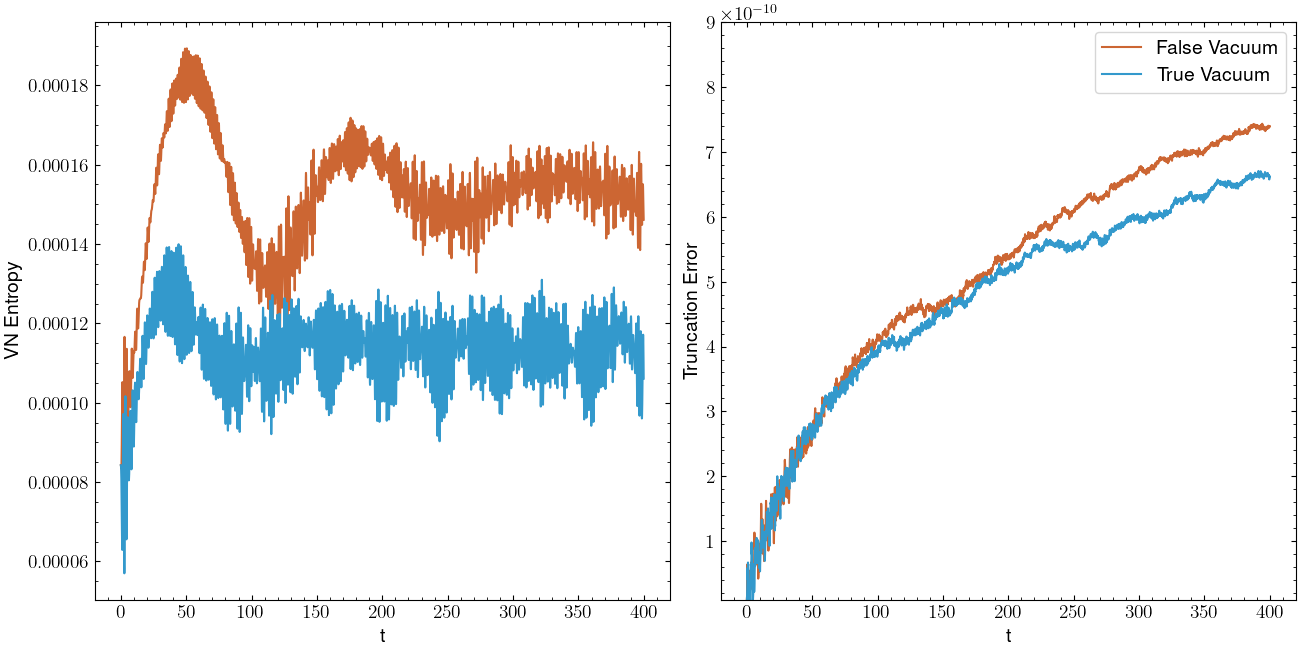}
    \caption{
    Von Neumann entropy (left) and Truncation Error (right) for the data employed in the main text. Parameters are $h_x = 0.2$, $h_z=0.04$ for the True Vacuum and $h_z=-0.04$ for the False Vacuum.}
    \label{fig:error}
\end{figure}

The numerical results shown in Fig. 2 in the main text were obtained using infinite Time Evolving Block Decimation (iTEBD) \cite{Vidal2007}. This method, based on infinite matrix product states, explicitly exploits translational invariance to simulate the dynamics on an infinite chain.
Concerning the specific quench we are performing, the entanglement is dimly growing for the bond dimension used in the simulation ($D=2048$): The Von Neumann entanglement entropy is stable — within the observation time — around $10^{-4}$ (Fig.~\ref{fig:error}) and the truncation error is consistently smaller than $10^{-9}$.

\begin{figure}
    \centering
    \includegraphics[width=\linewidth]{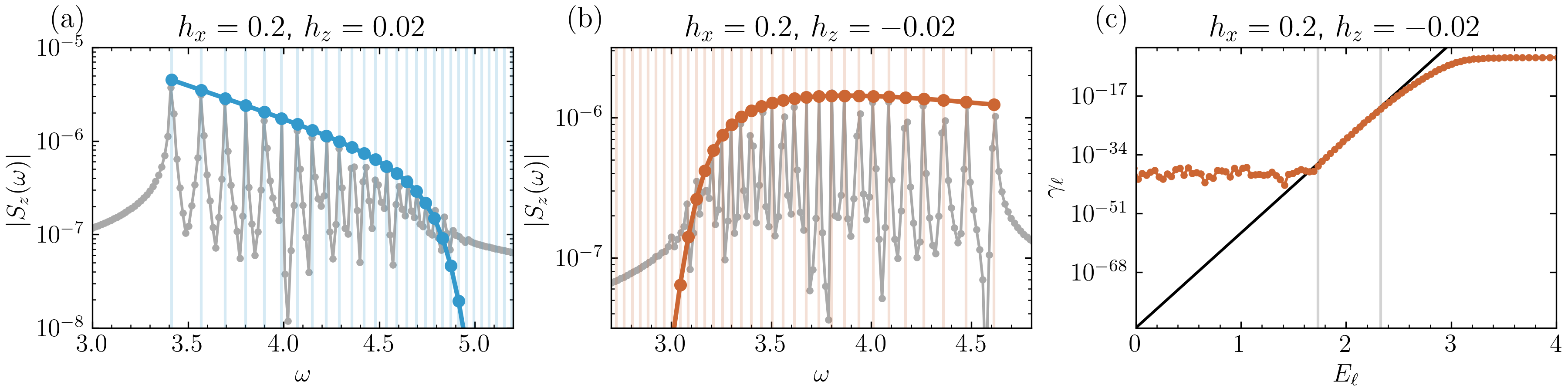}
    \caption{\label{fig3} Magnitude spectrum of the real-time evolved magnetization after a quench from (a) the true vacuum and (b) the false vacuum. Panel (c): exponential fitting of the amplitudes of the subcritical bubbles and $E_{\ell} \to 0$ extrapolation of the decay rate $\gamma$. Parameters are $h_x=0.2$ and  $|h_z|=0.02$. For a complete description of the different lines plotted here we refer to the caption of Fig.~2 in the main text.}
    \label{fig:fig2bis}
\end{figure}

While these small errors may appear surprising for such long evolution times, the reason can be easily understood: the energy density of the initial state with respect to the ground state of post-quench Hamiltonian is very small and the dynamics are mainly dominated by the bound oscillatory modes that we describe, generated at rest (with respect to the centre of mass momentum). At the same time, the emergence of propagating modes that could spread correlations across the system is largely suppressed \cite{Kormos2017}.
\\
In Fig.~\ref{fig:fig2bis} we show the results of the quench dynamics (similarly to Fig.~2 in the main text) for a different choice of parameters and we see similar agreement between the iTEBD simulation and the predictions based on the two-particle model.
\\

\section{Rydberg atom chain}
\label{sec:Rydberg}
We here show how the same quench protocol applied in the main text to the quantum Ising chain can also be used to probe the dynamical signatures of true vacuum bubbles in a Rydberg-atom experimental setup. We consider a one-dimensional array of optical tweezers, with one atom per tweezer. Each atom represents a qubit with two possible states: $\ket{g}$ is the ground state, and $\ket{r}$ is a Rydberg state. For each atom we define the (hard-core) bosonic operators $b_j = \ket{g}\bra{r}_j$, $b_j^\dagger = \ket{r}\bra{g}_j$ and $n_j=b_j^\dagger b_j$. The Hamiltonian governing the system is
\begin{equation}
    H_\text{Ryd}=\frac{\Omega}{2}\sum_j (b_j+b_j^\dagger)+\sum_j \delta_j n_j +\sum_{i,j} V_{i,j} n_i n_j,
\end{equation}
where $\Omega$ is the Rabi frequency, $\delta_j=\delta_0+(-1)^j \delta_\pi$ is a site-dependent detuning, and $V_{i,j}$ is the Rydberg-Rydberg interaction. We work in the {\it Rydberg blockade} regime, i.e., we assume that $V_{i,i+1}\gg \Omega, \delta_0, \delta_\pi$, such that we can project the dynamics in the constrained Hilbert space with $n_j n_{j+1}=0\; \forall j$. Interactions are expected to decay very fast with distance (as $1/r^6$), so we neglect interactions beyond nearest-neighbor. For $\delta_\pi=0$ and $\delta_0/|\Omega|<-0.655$, the ground-state spontaneously breaks translation symmetry \cite{Fendley2004,Rico2014}: there are two degenerate ground states (the {\it charge density waves}), that correspond to the configurations $\ket{\dots grgr\dots}, \ket{\dots rgrg\dots}$ in the limit $\Omega=0$. These two degenerate states represent the two vacua of the theory. If we then turn on $\delta_\pi\neq 0$, we explicitly break translation symmetry and the two charge density waves become a true and a false vacuum. Following our protocol, we use exact diagonalization to prepare the ground state of the Hamiltonian with $\delta_\pi=-10^{-3}$, and we then evolve it with $\delta_\pi> 0$ (false vacuum) or $\delta_\pi<0$ (true vacuum). 

\begin{figure}[h]
    \centering
    \includegraphics[width=0.49\columnwidth]{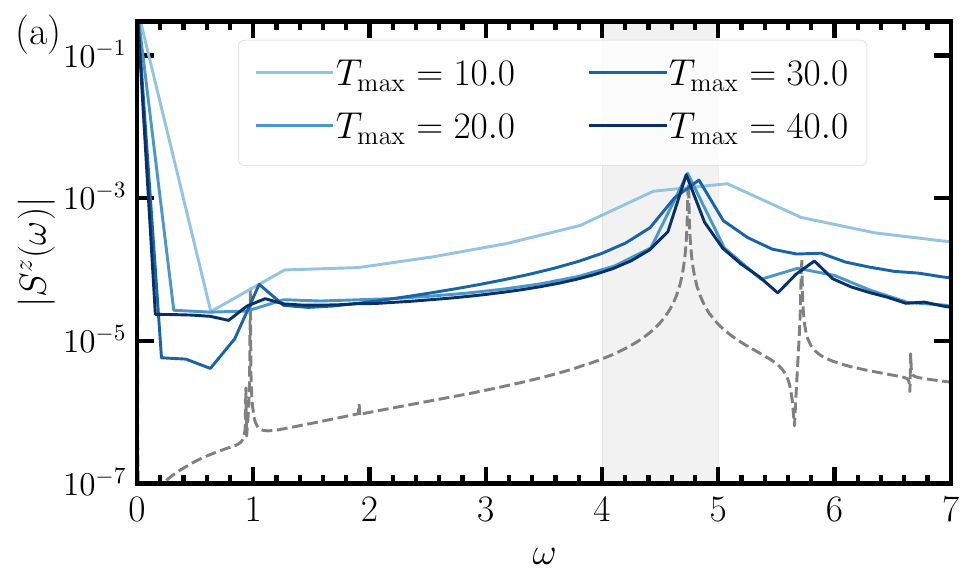}
    \includegraphics[width=0.49\columnwidth]{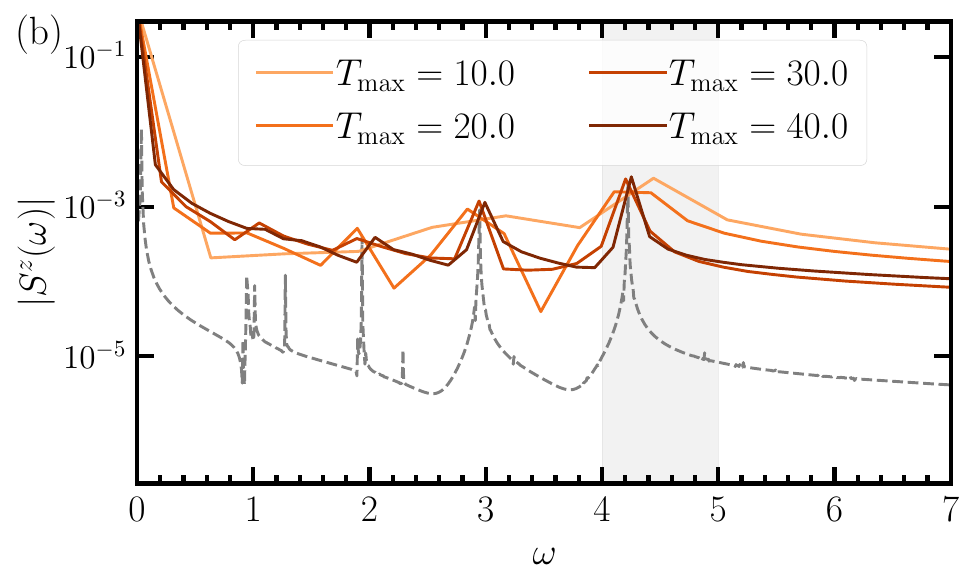}
    \caption{Fourier transform of the time evolution of the order parameter, for different total evolution times. The dashed gray line is obtained for a long evolution time $T_\mathrm{max}=1000$. The region near the maximum of the magnitude spectrum is indicated by a grey shading.}
    \label{fig:Ryd}
\end{figure}

In Fig.~\ref{fig:Ryd} we plot the Fourier transform of the oscillations of the order parameter (dashed gray line). In the numerical simulation we used $\Omega=1$, $\delta_0=4$, $|\delta_\pi|=0.5$, and a system size of $24$ sites with periodic boundary conditions. When the quench is performed from the true vacuum [Fig.~\ref{fig:Ryd}-(a)], we observe a large peak at $\omega \approx 4.7$, followed by two smaller peaks at $\omega \approx 5.7$ and $\omega \approx 6.7$. These peaks, whose amplitudes decrease with increasing frequency, have a similar profile to the ones observed in Fig.~2-(a,d) in the main text. An additional peak at $\omega\approx 1$ corresponds to the difference between the frequencies of the mesons. Also for the false vacuum [Fig.~\ref{fig:Ryd}-(b)], we observe a sequence of peaks that is very similar to the one shown in Fig.~2-(b,e) in the main text: largest frequency is $\omega\sim 4.2$, and both the spacings between the peaks and the amplitudes decrease as $\omega$ decreases. 

As explained in the main text (Fig. 1), the difference between the true and false vacuum can be observed by analyzing the peaks near the maximum of the magnitude spectrum (indicated here by a shaded region, with $\omega_\text{max}\approx 4.5$). In the region $\omega<\omega_\text{max}$, for the false vacuum we observe a sequence of peaks with amplitudes that decrease away from $\omega_\text{max}$. On the other hand, the true vacuum's magnitude spectrum does not exhibit such a sequence of peaks, and rapidly drops for $\omega<\omega_\text{max}$. 

In order to observe the peaks in an experiment, a long evolution time may be needed to achieve the desired frequency resolution. Since the experiment has limited coherence time, it is crucial to understand whether a moderate evolution time is enough to characterize the oscillations. In Fig. \ref{fig:Ryd} we see that the peaks become clearly visible for an evolution time $T_\mathrm{max}\sim 20$.

\bibliography{bib}